\documentclass[12pt]{article}

\ifx\pdfoutput\undefined
\usepackage[dvips,bookmarks]{hyperref}  
\else
\usepackage{hyperref}   
\fi

\usepackage{graphicx}

\hypersetup{colorlinks,bookmarksopen,bookmarksnumbered,citecolor=blue,
pdfstartview=FitH}
\def\hhref#1{\href{http://arxiv.org/abs/hep-th/#1}{hep-th/#1}} 
\def\mhref#1{\href{mailto:#1}{#1}}              

\def\bop#1{\setbox0=\hbox{$#1M$}\mkern1.5mu
\vbox{\hrule height0pt depth.04\ht0
\hbox{\vrule width.04\ht0 height.9\ht0 \kern.9\ht0
\vrule width.04\ht0}\hrule height.04\ht0}\mkern1.5mu}
\def\bo{{\mathpalette\bop{}}}                        
\def\frac#1#2{{\textstyle{#1\over#2}}}     

\parskip=\medskipamount         
\thicklines             
\begin{document}

\newcommand{\be}{\begin{equation}}
\newcommand{\ee}{\end{equation}}
\newcommand{\mx}{\mbox}
\newcommand{\mt}{\mathtt}
\newcommand{\p}{\partial}
\newcommand{\st}{\stackrel}
\newcommand{\al}{\alpha}
\newcommand{\bb}{\beta}
\newcommand{\ga}{\gamma}
\newcommand{\te}{\theta}
\newcommand{\de}{\delta}
\newcommand{\et}{\eta}
\newcommand{\ze}{\zeta}
\newcommand{\s}{\sigma}
\newcommand{\e}{\epsilon}
\newcommand{\om}{\omega}
\newcommand{\Om}{\Omega}
\newcommand{\la}{\lambda}
\newcommand{\La}{\Lambda}
\newcommand{\ti}{\widetilde}
\newcommand{\ih}{\hat{i}}
\newcommand{\jh}{\hat{j}}
\newcommand{\kh}{\widehat{k}}
\newcommand{\lh}{\widehat{l}}
\newcommand{\eh}{\widehat{e}}
\newcommand{\ph}{\widehat{p}}
\newcommand{\qh}{\widehat{q}}
\newcommand{\mh}{\widehat{m}}
\newcommand{\nh}{\widehat{n}}
\newcommand{\Dh}{\widehat{D}}
\newcommand{\2}{{\textstyle{1\over 2}}}
\newcommand{\3}{{\textstyle{1\over 3}}}
\newcommand{\4}{{\textstyle{1\over 4}}}
\newcommand{\8}{{\textstyle{1\over 8}}}
\newcommand{\6}{{\textstyle{1\over 16}}}
\newcommand{\ra}{\rightarrow}
\newcommand{\lra}{\longrightarrow}
\newcommand{\Ra}{\Rightarrow}
\newcommand{\im}{\Longleftrightarrow}
\newcommand{\hs}{\hspace{5mm}}
\newcommand{\bea}{\begin{eqnarray}}
\newcommand{\eea}{\end{eqnarray}}
\newcommand{\NP}{{\em Nucl.\ Phys.\ }}
\newcommand{\AP}{{\em Ann.\ Phys.\ }}
\newcommand{\PL}{{\em Phys.\ Lett.\ }}
\newcommand{\PR}{{\em Phys.\ Rev.\ }}
\newcommand{\PRL}{{\em Phys.\ Rev.\ Lett.\ }}
\newcommand{\PRP}{{\em Phys.\ Rep.\ }}
\newcommand{\CMP}{{\em Comm.\ Math.\ Phys.\ }}
\newcommand{\MPL}{{\em Mod.\ Phys.\ Lett.\ }}
\newcommand{\IJMP}{{\em Int.\ J.\ Mod.\ Phys.\ }}

\begin{titlepage}
\setcounter{page}{0} 
\begin{flushright}
YITP-SB-06-59\\ December 14, 2006\\
\end{flushright}

\begin{center}
{\centering {\LARGE\bf
String interactions on the random lattice
\par} }

\vskip 1cm {\bf Haidong Feng\footnote{E-mail address:
\mhref{hfeng@ic.sunysb.edu}}, Yu-tin Huang\footnote{E-mail address:
\mhref{yhuang@grad.physics.sunysb.edu}}, Warren Siegel\footnote{E-mail
address: \mhref{siegel@insti.physics.sunysb.edu}}} \\ \vskip 0.5cm

{\it C.N. Yang Institute for Theoretical Physics,\\
State University of New York, Stony Brook, 11790-3840 \\}

\end{center}

\begin{abstract}
We combine two partons on a random lattice as a vector state.
In the ladder approximation, we find that such states have $1/p^2$
propagators (after tuning the mass to vanish). We also construct some diagrams
which are very similar to 3-string vertices
in string field theory for the first oscillator mode. Attaching 3 such lattice states to these
vertices, we get Yang-Mills and $F^3$ interactions
up to 3-point as from bosonic string (field)
theory. This gives another view of a gauge field as a
bound state in a theory whose only fundamental
fields are scalars.
\end{abstract}

\end{titlepage}

\section{Introduction}\label{1}

It is known that, in nonrelativistic quantum mechanics, Regge behavior
relates the angular momenta and energies of
bound states \cite{Regge}. In relativistic quantum field theory,
the high-energy behavior of a scattering amplitude, $F(s,t) \sim  \beta (s)
t^{\alpha (s)}$ as $t \to \infty$ and $s < 0$, is also dominated by
Regge poles, with trajectories $J= \alpha (s)$. Here the
Bethe-Salpeter equation \cite{bethe} takes the place of the Schr\"odinger
equation, which can only be solved in certain approximations, such as the ladder
approximation or a perturbative Feynman diagram analysis.

Experimental data confirms the existence of families of particles
along trajectories $J= \alpha(s)$ which are {\em linear} as from the Veneziano model or
string (field) theory. However, in many approximations of
conventional field theory the trajectories rise for a while
and then fall back towards negative values of $J$ for increasing energy. Thus,
only a few bound states are produced, as characteristic of a Higgs phase;
instead, linearity and an infinite number of bound states are expected to arise
as a consequence of confinement, perhaps due to some infrared catastrophe.
However, such a catastrophe is absent in the usual calculations, which are always
made for massive or off-shell states precisely in order to avoid infrared
divergences.

Originally, strings were introduced for hadrons and later identified as
bound states of ``partons". Unfortunately, a suitable hadronic
string theory serving that purpose hasn't been constructed. This led
to the reinterpretation of the known strings as fundamental strings
describing gluons and quarks, leptons, gravitons, etc. The target space
is 26D for the bosonic theory and 10D for the super theory, which means
compactification is necessary.

One nonperturbative approach to strings is quantization on a suitable random lattice
representing the worldsheet \cite{random}. It expresses the strings as bound
states of underlying partons, and the lattices are
identified with Feynman diagrams \cite{David}. The two theories are ``dual" to each
other, and one is perturbative while the other is nonperturbative.
The Feynman diagrams of the particles underlying
this bosonic string were studied and linear Regge
trajectories were reproduced in the ladder approximation \cite{ladders}.
This implies that the only fundamental
fields are scalars and all others can be represented as composite fields.
Here we will show an example how the gauge field can be constructed
as a composite field of partons.

The outline of this article is: section \ref{2}, a brief review of Reggy theory;
section \ref{3}, a review of the bosonic lattice string; section \ref{4}, reproducing
linear Regge trajectories in the ladder approximation; section \ref{5}, introducing
the new massless external state from the lattice; section \ref{6}, constructing
two simple 3-state interacting diagrams on the lattice, computing the 3-point interactions
similar to the usual Yang-Mills field; section \ref{7}, comparing the two 3-state vertices
with Witten's vertex and the Caneschi-Schwimmer-Veneziano (CSV) vertex in string field theory; and section \ref{8}, discussions.

\section{Regge theory}\label{2}

As suggested by Regge, Regge poles might be
relevant to the analysis of high-energy scattering.
Many results about poles' locations and properties were
obtained on the basis of analyticity assumptions, mostly in
$\phi^3$ theory \cite{eden}. A simpler consideration
is to examine the high-energy behavior of scattering amplitudes
directly by summing suitable sets of Feynman diagrams \cite{federbush}.

The two-particle elastic scattering amplitude $A(s,t)$ for
an appropriate set of Feynman diagrams (e.g., ladders) can be of the form:
\be
A(s,t) = \int d^4k_i \prod_a\frac{1}{{p_a^2 +m^2}} \sim \int  d^4k_i \int_0^{\infty}
\prod_a d\beta_a e^{-\beta_a (p_a^2 +m^2)/2}
\ee
where $k_i$ are independent loop momenta, $\beta_a$ are Schwinger
parameters to exponentiate the propagators and the Mandelstam variables are
\be s = -(q_1+q_2)^2= -(q_3+q_4)^2, \quad t = -(q_1-q_3)^2 \ee
As we will see, the only
difference between an ordinary field theory and lattice string theory
is the integration
over the parameters $\beta_a$. In the lattice string case reviewed
in the next section, they are fixed at $\beta_a=\alpha '$.

Integrating out Gaussian loop momenta,
\be A(s,t) \sim \int_0^{\infty} \prod_a d\beta_a
\frac{N(\beta)}{[C(\beta)]^2} e^{-g(\beta)t -d(s, \beta)}
\ee
When $t \to \infty$, it is
dominated by the region near $g(\beta) =0$.
So to make the coefficient of $t$
vanish, one can set those $\beta$'s to zero everywhere except in $g(\beta)$,
which shortcircuits the diagram to eliminate the $t$ dependence. Then the integration
can be carried out to obtain the asymptotic behavior as $t \rightarrow 0$. For ladder graphs,
the ladder with $n$ rungs has an expression of the form:
\be
 A_n(s,t) \sim
g^2 \frac{1}{t} [g^2K(s) \ln t]^{n-1}
\ee
where $K(s)$ is just a self-energy
diagram evaluated from a bubble in 2 fewer dimensions. So the asymptotic behavior
comes from the sum of ladder diagrams:
\be
 \sum A_n(s,t)/(n-1)! = g^2 t^{\alpha (s)},~~~~ \alpha(s) = -1 +g^2K(s)
\ee
which is the result associated with the Regge trajectory.

\section{Bosonic lattice string review}\label{3}

The main difference between the lattice and continuum approaches to the string is that a
lattice requires a scale, while conformal invariance of the continuum string
includes scale invariance. To break the conformal invariance of the worldsheet,
a term proportional to the area (the simplest scale-{\it variant} and coordinate-{\it invariant} property
of the worldsheet) with coefficient (cosmological constant) $\mu$ is added to the
string action. Furthermore, to describe the string interaction, the string coupling
constant, which is counted by the integral of the worldsheet
curvature $R$, should be included. So totally, the action is
\begin{equation}
S = \oint {d^2 \sigma\over 2\pi} \sqrt{-g} \left[\frac{1}{\alpha'}
g^{mn} \frac{1}{2} (\partial_m X \cdot \partial_n X) + \mu + (\ln
\kappa) \frac{1}{2} R \right]
\end{equation}
On the random lattice, this action can be written as
\begin{equation}\label{action}
S_1 = \frac{1}{\tilde{\alpha}'} \sum_{\langle ij \rangle} \frac{1}{2} (x_i - x_j)^2
+ \mu \sum_i 1 + \ln \kappa \left(\sum_i 1 - \sum_{\langle ij \rangle}
+ \sum_J 1\right)
\end{equation}
where $j$ are vertices, $\langle ij \rangle$ the links (edges), and $J$ the
plaquets (faces, planar loops) of the lattice. The functional integration
over the worldsheet metric in usual string theory is repalced by a sum over
Feynman diagrams. The positions of vertices are integrated (except external
vertices; alternatively, external states will be introduced to calculate the
full amplitudes, as shown in later sections):
\be A = \sum \int \prod dx\ e^{-S_1}= \sum e^{-\mu \sum_i 1}\int  dx\ \prod_{ij}
e^{-\frac{1}{2 \widetilde{\alpha}'}  (x_i-x_j)^2 }    \ee

Now, by identifying the lattice with a position-space Feynman diagram,
we can find the underlying field theory as follows:
Vertices of the lattice correspond to those of Feynman diagram and links to
propagators; the 1/N expansion is associated with the faces of the worldsheet polyhedra
with $U(N)$ indices. Thus, the area term (counting the number of vertices) in the lattice
action (\ref{action}) gives the coupling constant factor for each vertex in the field
action, and the worldsheet curvature term gives the string coupling 1/N of the topological
expansion \cite{Hooft}. Explicitly, the action of an $n$-point-interaction scalar-field action is
\begin{equation}
S_2 = N\ tr \int {d^D x\over (2\pi \widetilde{\alpha}')^{D/2}}
\left(\frac{1}{2} \phi e^{-\widetilde{\alpha}' \bo /2} \phi - G \frac{1}{n} \phi^n \right)
\end{equation}
with
\begin{equation}
G = e^{-\mu} , \quad \quad \frac{1}{N} = \kappa
\end{equation}
The interaction $\phi^n$ can be chosen arbitrarily; restrictions may come
from consistency of the worldsheet continuum limit \cite{W. Siegel1}. In this paper, we will focus
on the minimum coupled lattice, $\phi^3$ theory, but the calculation
for a $\phi^4$ interaction is pretty much the same.

\section{Ladder graphs and Regge trajectories}\label{4}

In this section we review the ladder graphs responsible for
a Regge trajectory $\alpha(s)$, and compare with those done in the early days
of Regge theory. Since somewhat similar procedures will be used in following sections, we give
details in this section.

\begin{figure}[ht]
\begin{center}
\includegraphics[scale=0.5]{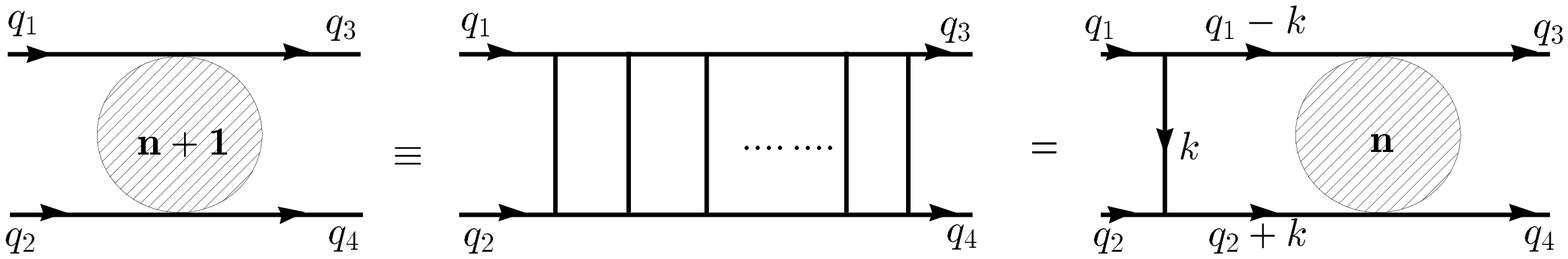}
\end{center}
\caption{\label{fig1} Ladder diagrams}
\end{figure}

Consider 4-point functions in the parton theory with Gaussian propagators and
cubic interaction $\phi^3$ with coupling constant $\lambda$.
The amplitude is evaluated by solving the Bethe-Salpeter
equation in the ladder approximation with two incoming particles of momenta $q_1$ and $q_2$
and two outgoing particles of momenta $q_3$ and $q_4$, as depicted in Fig.~\ref{fig1}.

The two-particle propagator $\Delta$ satisfies the Bethe-Salpeter equation in $D$ dimentions
\begin{equation}\label{BS}
\Delta = 1 + e^{-H} \Delta
\end{equation}
where $e^{-H}$ sticks an extra rung on the sum of ladders  (as in Fig.\ 1).
Explicitly, it can be written as
\be\label{propagator}  e^{-H} = (\hbox{rung propagator})\times(\hbox{two ``side" propagators}) \ee
with integration over either loop momentum or positions of vertices. The propagator is given by
\begin{equation}
\Delta = {1\over 1 -e^{-H}}= \sum(e^{-H})^n
\end{equation}
Here, we will replace integrals with operator expressions as in usual string theory.
Thus, adding the two sides followed by adding the rung in (\ref{propagator})
is performed by the operator
\be\label{e^H} e^{-H} = e^{-(x_1-x_2)^2/2}e^{-(p_1^2+p_2^2)/2} \ee
where the $p$'s and $x$'s are now the operators for the two particles.
Separating $p$'s and $x$'s into average and relative coordinates,
\begin{equation}
p_{1,2} = P \pm p , \quad x_{1,2} = \frac{1}{2} X \pm \frac{1}{2} x
\end{equation}
(\ref{e^H}) is then
\be e^{-H} = e^{-x^2/2}e^{-P^2+p^2} = e^{-x^2/2}e^{-p^2}e^{s/4} \ee
where
$$P^2= -\frac{1}{4} (q_1+q_2)^2= -\frac{1}{4} (q_3+q_4)^2=-s/4$$
By a similarity transformation, we can put half of
one exponential on each side,
\be\label{similarity transform} e^{-H} \quad\to\quad  e^{s/4}e^{-x^2/4}e^{-p^2}e^{-x^2/4}
\quad or \quad e^{s/4}e^{-p^2/2}e^{-x^2/2}e^{-p^2/2} \label{herm}\ee
To write $H$ as a manifestly Hermitian expression, we apply the Baker-Campbell-Haussdorf theorem to
combine the exponentials into a single one. Because the exponents,
$\frac{1}{2} x^2, \frac{1}{2} p^2$, satisfy the commutation relations of
raising and lowering operators and the Baker-Campbell-Haussdorf theorem
requires only commutators, we can use the representation

\be \2 x^2 \to \pmatrix{ 0 & 1 \cr 0 & 0 \cr}, \quad
\2 p^2 \to \pmatrix{ 0 & 0 \cr 1 & 0 \cr}, \quad i \2 \{x,p\} \to \pmatrix{ 1 &
0 \cr 0 & -1 \cr} \ee
So, in general,
{\jot=.2in
\bea
&& e^{-\alpha p^2/2}e^{-\beta x^2/2}e^{-\alpha p^2/2}\quad\to\quad
e^{-\left({0\atop\alpha}{0\atop 0}\right)}e^{-\left({0\atop 0}{\beta\atop
0}\right)} e^{-\left({0\atop\alpha}{0\atop 0}\right)} \nonumber \\ & = &
\pmatrix{ 1 & 0 \cr -\alpha & 1 \cr}\pmatrix{ 1 & -\beta \cr 0 & 1 \cr}
\pmatrix{ 1 & 0 \cr -\alpha & 1 \cr} = \pmatrix{ 1+\alpha\beta & -\beta \cr
-\alpha(2+\alpha\beta) & 1+\alpha\beta \cr} \nonumber \\  & = & e^{-\left({0\atop b}{a\atop
0}\right)} =  cosh(\sqrt{ab}) - {sinh(\sqrt{ab})\over
\sqrt{ab}}\pmatrix{ 0 & a \cr b & 0 \cr}
\eea
}
Then, $H$ in (\ref{similarity transform}) becomes the hamiltonian of a harmonic oscillator
\begin{eqnarray}\label{H}
H & = & -\frac{1}{4} s - ln \lambda^2 + \omega (m \omega \frac{1}{2} x^2 + \frac{1}{m\omega}
\frac{1}{2} p^2) \nonumber \\
& = & -\frac{1}{4} s - ln \lambda^2 + \frac{\omega}{2} D + \omega a^{\dagger} \cdot a
\end{eqnarray}
with $\lambda$ restored. If we work in coordinate space, as in the following sections,
\be \alpha = \frac{1}{2} , \beta = 2 \quad\Rightarrow\quad
\omega = ln(2+\sqrt 3) , \quad m\omega = {\sqrt 3 \over 2} \ee
we can find the Regge trajectory from the spectrum of this harmonic oscillator.
The harmonic oscillators (a $D$-vector) can be interpreted as the oscillators
in the usual string theory (but only one such vector) as follows: The
positions of the two partons in the Bethe-Salpeter equation are two adjacent
points on the random lattice, and the relative coordinate represents the
first order derivative of $x(\sigma)$ corresponding to the first oscillator.
(A similar model was considered in \cite{susskind}.)

Taking $(D/2)\omega$ as the ground-state energy and integer
excitation $J$ as the (maximum) spin of the $D$
oscillators (acting with $J$ vector oscillators on the vacuum),
the ``energy" of the harmonic oscillator Hamiltonian
$m\omega^2\2 x^2 +(1/m)\2 p^2$ can be identified as $(J+D/2)\omega$.
Since the Bethe-Salpeter equation corresponds to perturbatively solving a Schr\"{o}dinger
equation with ¡°free¡± Hamiltonian 1, potential $e^{-H}$ and vanishing total energy $e^{-H}-1=0$,
it gives $H=2 \pi i n$,
\be 2\pi in = -\4 s -ln(\lambda^2) +\omega (J+\2 D) \ee
So we have the trajectory $J= \alpha (s)$
\be\label{trajectory} \alpha(s) = -\2 D +{1\over \omega}
[\4 s +ln(\lambda^2) +2\pi in] \ee
The real part of (\ref{trajectory})
\be\label{realtrajectory} \alpha(s) = -\2 D +{1\over \omega}
[\4 s +ln(\lambda^2) ] \ee
is linear with positive slope. The real pole gave us the asymptotic behavior,
while complex poles do not affect the Regge trajectory, as shown in \cite{ladders}.
We require the vertical intercept of this Regge trajectory, which is given by $s=0$, to be $J=1$, so the corresponding spin-one particle is massless. Thus $\alpha(0) =1$ gives
\be\label{intercept} e^{- \omega(D+2)} \lambda^4 =1 \ee
(In the usual continuum approach, this constraint, as well as $D=26$, are found perturbatively, but in the lattice approach they would be nonperturbative, so we impose them by hand.)

There are several ways to interpret the group theory of this state:  (1) We can examine only color-singlet states (the partons are N by N matrices of U(N) color); then we should take the color trace of this vector, which would make it Abelian.  (2) If we examine color-nonsinglets, the vector is in the adjoint representation, and so represents a Reggeized bound-state gauge field of color, and thus not a true string state.  (3) If we introduce a second type of scalar parton which is in the fundamental representation of both color and a second, ``flavor" symmetry, we can consider ladders where these scalar ``quarks" run along the outside, giving an open string instead of a closed one \cite{Hooft}.  Then the vector is the gauge field of this flavor symmetry.  It is really only in this last case that string theory implies the state is massless.

\section{External vertex operator for gauge field}\label{5}

Now we are ready to introduce the ground state and first excited state for the harmonic oscillator in ladders (\ref{H}):
$$ e^{ik \cdot (x_i +x_j)/2} | 0 \rangle \quad and \quad \epsilon \cdot (x_i - x_j) e^{ik \cdot (x_i +x_j)/2} | 0 \rangle$$
They are the very same as the vertex operators $e^{ik \cdot X}$ and  $\epsilon \cdot \partial X e^{ik \cdot X}$ in the bosonic string, except latticized.

Defining
$$x= x_i - x_j =
\sqrt{\frac{1}{2m\omega}}(a + a^{\dagger})$$
where $a, a^{\dagger}$ are
creation and annihilation operators of the harmonic oscillator in ladders, the first excitation can also be written as
\be\label{gaugestate} \sqrt{\frac{1}{2m\omega}} \epsilon \cdot a^{\dagger} | 0 , k \rangle
\ee
with
\be | 0 , k \rangle = e^{ik \cdot (x_i +x_j)/2} | 0 \rangle\ee
As in usual string or string field theory, this first excited state should be a massless state and the propagator should have a massless pole. To check it, let's consider the amplitude for one
incoming and one outgoing state with momenta $k$ and $k'$
respectively. In the ladder approximation as reviewed in the last section, we have to
evaluate the amplitude depicted in Fig.~\ref{fig1}:
\begin{equation}\label{2point}
{\cal A} = - \frac{1}{2m\omega} \langle 0, k | (\epsilon_1 \cdot a) \Delta
(\epsilon_2 \cdot a^{\dagger}) | 0 , k' \rangle
\end{equation}
with the definitions $k= q_1 + q_2$ and $k' = q_3 +q_4$.
The calculation is pretty similar to the previous section. The
two-particle propagator $\Delta= \frac{1}{1- e^{-H}}$ satisfying the Bethe-Salpeter
equation as in (\ref{BS}) and $H$ is expressed by annihilation (creation) operators
as in (\ref{H}).

In such ladder approximations, the propagator should be written as a summation of all ladders
\begin{equation}
\Delta = \frac{1}{1- e^{-H}} = \sum (e^{-H})^n
\end{equation}
Thus, using the commutator $[a_{\mu}, a^{\dagger}_{\nu} ] = \delta_{\mu , \nu}$ and integrating out
$X$'s, the amplitude in (\ref{2point}) is
\begin{eqnarray}\label{2point 1}
{\cal A} & = & - \frac{1}{2m\omega} \langle 0 ,k | (\epsilon_1 \cdot a) \frac{1}{1- e^{-H}}
(\epsilon_2 \cdot a^{\dagger}) | 0 , k' \rangle \nonumber \\
& = & - \frac{1}{2m\omega} \frac{\epsilon_1 \cdot \epsilon_2}{1-e^{s/4}
\lambda^2 e^{-\omega (1+D/2)}} \delta^D (k+k')
\end{eqnarray}
As given in (\ref{intercept}), the real Regge trajectory
$\alpha(0) =1$ gives $e^{- \omega(D+2)} \lambda^4 =1$.
Then (\ref{2point 1})
\begin{eqnarray}
{\cal A} & = & - \frac{1}{2m\omega} \frac{\epsilon_1 \cdot \epsilon_2}{1-e^{-k^2/4}} \delta^D (k+k') \\
& = & - \frac{2}{m\omega} \frac{1}{k^2} \epsilon_1 \cdot \epsilon_2 \quad , \quad k^2 \rightarrow 0
\end{eqnarray}
has a massless pole.

This result can also be seen from the ladder integration if we rewrite (\ref{2point}) as
\begin{eqnarray}\label{2point 2}
{\cal A} & = & - \frac{1}{2m\omega} \sum_n \langle 0 ,k | (\epsilon_1 \cdot a) (e^{-H})^n
(\epsilon_2 \cdot a^{\dagger}) | 0 , k' \rangle \nonumber \\
& = & - \frac{1}{2m\omega} \sum_n A_n
\end{eqnarray}
Here $A_n$ is the amplitude for a single ladder with $n$ loops (including external loops)
\begin{eqnarray}\label{n-ladder}
A_n &=& \int (\prod_{i=0}^{n} d^D x_i d^D y_i)\langle 0 , k |(\epsilon_1 \cdot a)|
x_0, y_0 \rangle  [\prod_{i=1}^{n} (e^{-H_i})] \langle x_n, y_n |(\epsilon_2 \cdot a^{\dagger})
| 0,k' \rangle
\end{eqnarray}
with
\begin{eqnarray}\label{externstate}
&& \langle 0 , k |(\epsilon_1 \cdot a) | x_0, y_0 \rangle = \epsilon_1 \cdot (x_0 - y_0)
e^{-\frac{m\omega}{2}(x_0 - y_0)^2} e^{ik \cdot \frac{(x_0 + y_0)}{2}} \nonumber \\
&& \langle x_n, y_n |(\epsilon_2 \cdot a^{\dagger}) | 0, k' \rangle = \epsilon_2 \cdot (x_n - y_n)
e^{-\frac{m\omega}{2}(x_n - y_n)^2} e^{ik' \cdot \frac{(x_n + y_n)}{2}}
\end{eqnarray}
according to the definition of the ground state of the harmonic oscillator, and
\begin{eqnarray}
H_i & = & \langle x_{i-1}, y_{i-1} |H | x_i, y_i \rangle \nonumber \\
& = & \ln(\lambda^{-2}) + \frac{1}{4} s + \frac{1}{4} (x_{i-1} - y_{i-1})^2
+ \frac{1}{2} (x_{i-1} - x_i)^2 \nonumber \\ && + \frac{1}{2} (y_{i-1} - y_i)^2
+\frac{1}{4} (x_i - y_i)^2
\end{eqnarray}
Doing the Gaussian integrals for $x_i$'s and $y_i$'s,  (\ref{n-ladder}) becomes
\begin{eqnarray}
A_n = -\frac{\epsilon_1 \cdot \epsilon_2}{2m\omega} [\lambda^2 e^{-k^2/4} e^{-\omega (1+D/2)}]^n \delta^D (k+k')
\end{eqnarray}
which gives the same massless pole as in (\ref{2point 1}).

This massless pole means the first excited state (\ref{gaugestate}) is a massless state, and has the same propagator as the YM gauge field in Feynman gauge. We will discuss in the following sections how this generalizes to interactions.

\section{The 3-string vertex}\label{6}

To get the 3-point gauge interaction in YM fields, we have to find a way to join three states.
One analogue is Witten's open string field theory, in which stings interact
by identifying the right half of each string with the left half of the next one. On the lattice we need to sum over an infinite number of diagrams representing this situation, each giving a result very similar to the 3-vector vertex in string field theory.  Here we will give the 2 simplest examples to show that they give the same interaction as the usual YM field.

Similarly to string field theory (SFT), the 3-state interaction can be written as
\begin{equation}\label{A3}
{\cal A}^{(0)}_3 = (\langle 0|_1 \otimes
\langle 0 |_2 \otimes \langle 0 |_3 ) | V_1 V_2 V_3 \widehat{F} |0 \rangle
\end{equation}
Using the definitions
\begin{eqnarray}
x_1 + x_2 = X \quad &,& \quad x_1 - x_2 = x ; \nonumber \\
y_1 + y_2 = Y \quad &,& \quad y_1 - y_2 = y ; \nonumber \\
z_1 + z_2 = Z \quad &,& \quad z_1 - z_2 = z ;
\end{eqnarray}
$$V_1 = \epsilon_1 \cdot x e^{ik \cdot X/2},\quad V_2 = \epsilon_2 \cdot y e^{ik \cdot Y/2} \quad and\quad V_3 = \epsilon_3 \cdot z e^{ik \cdot Y/2}$$
are the external vertex operators for massless fields as considered in the previous section. In the operator formulation,
$$x=\sqrt{\frac{1}{2m\omega}}(a + a^{\dagger}), \quad y=\sqrt{\frac{1}{2m\omega}}(b + b^{\dagger}),\quad
z=\sqrt{\frac{1}{2m\omega}}(c + c^{\dagger})$$
where $a(a^{\dagger})$, $b(b^{\dagger})$, $c(c^{\dagger})$ are annihilation (creation) operators for three independent harmonic oscillators in the ladder approximation.

\begin{figure}[ht]
\begin{center}
\includegraphics[scale=0.5]{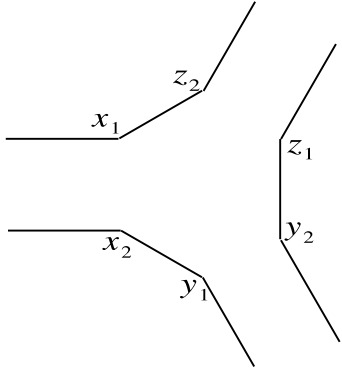}
\end{center}
\caption{\label{fig2}  The interaction lattice of order $\lambda^0$ with
the vertex given by $\widehat{F}$.}
\end{figure}

The simplest figure for the 3-string lattice vertex is shown in Fig.~\ref{fig2}.
Then the 3-state vertex can be constructed with annihilation (creation) operators of ladders after integrating out $X,Y,Z$
\begin{eqnarray}\label{V30}
\widehat{F} &=& \int d^D X d^D Y d^D Z e^{-\frac{1}{2} [x_1-z_2)^2 + (y_1 - x_2)^2 + (z_1 - y_2)^2]
+ \frac{i}{2} (k_1 \cdot X + k_2 \cdot Y + k_3 \cdot Z)} \nonumber \\
&=&e^{-\frac{1}{6}(x+y+z)^2 -\frac{i}{6}(x \cdot k_{23} + y \cdot k_{31} + z \cdot k_{12} )}
\end{eqnarray}
where $x,y,z$ can be expressed with annihilation (creation) operators $a(a^{\dagger})$,
$b(b^{\dagger})$,$c(c^{\dagger})$ and $k_{ij} = k_i -k_j$.

Thus the amplitude in (\ref{A3}) can be evaluated using commutators of 3 annihilation (creation) operators and the Baker-Campbell-Haussdorf theorem. First we write
\begin{equation}
-\frac{1}{6} (x+y+z)^2 = -\frac{1}{6} \frac{1}{2m\omega} [(a^{\dagger} + b^{\dagger} +
c^{\dagger})^2 + (a+b+c)^2 + \{a+b+c, a^{\dagger}+b^{\dagger}+c^{\dagger}\}]
\end{equation}
It is noticed that the ingredients of exponents satisfy the commutation relations of
raising and lowering operators of SU(1,1). We use the representation
\begin{eqnarray}\label{split1}
& \frac{i}{6} (a+b+c)^2 \rightarrow \left(
\begin{array}{cc}
0 & 1 \\
0 & 0 \\
\end{array}
\right), \quad
\frac{i}{6} (a^{\dagger} + b^{\dagger}+ c^{\dagger})^2 \rightarrow \left(
\begin{array}{cc}
0 & 0 \\
1 & 0 \\
\end{array}
\right) , \nonumber \\
&-\frac{1}{6} \{a+b+c, a^{\dagger}+b^{\dagger}+c^{\dagger}\} \rightarrow
\left(
\begin{array}{cc}
1 & 0 \\
0 & -1 \\
\end{array}
\right)
\end{eqnarray}
to calculate the commutators when we apply the Baker-Campbell-Haussdorf theorem. Then
{\jot=.2in
\begin{eqnarray}
e^{-\frac{1}{6} (x+y+z)^2} &=& e^{\left({\frac{1}{2m\omega}\atop\frac{1}{2m\omega}}
{\frac{i}{2m\omega}\atop \frac{-1}{2m\omega}}\right)}
= \left(
\begin{array}{cc}
1+ \frac{1}{2m\omega} & \frac{i}{2m\omega} \\
\frac{i}{2m\omega} & 1-\frac{1}{2m\omega} \\
\end{array}
\right)  = e^{\left({0\atop\alpha}{0\atop 0}\right)} e^{-\left({\beta \atop 0}{0\atop -\beta}\right)}
e^{-\left({0\atop 0}{\alpha \atop 0}\right)} \nonumber \\
&=& \pmatrix{ 1 & 0 \cr \alpha & 1 \cr} \pmatrix{ e^{\beta} & 0 \cr 0 & e^{-\beta} \cr}
\pmatrix{ 1 & \alpha \cr 0 & 1 \cr}  = \pmatrix{ e^{\beta} & \alpha e^{\beta} \cr \alpha e^{\beta} &
\alpha^2 e^{\beta} + e^{-\beta} \cr} ,
\end{eqnarray}
}
which gives
\begin{equation}
\alpha = \frac{i}{1+2m\omega} , \quad \beta = \ln (1+ \frac{1}{2m\omega})
\end{equation}
Thus,
\begin{eqnarray}
e^{-\frac{1}{6} (x+y+z)^2} |0\rangle &=& e^{\alpha \frac{i}{6} (a^{\dagger} + b^{\dagger}+ c^{\dagger})^2 }
e^{-\frac{\beta}{6} \{a+b+c, a^{\dagger}+b^{\dagger}+c^{\dagger}\}} e^{\alpha \frac{i}{6} (a+b+c)^2} |0\rangle
\nonumber \\ &=& e^{\alpha \frac{i}{6} (a^{\dagger} + b^{\dagger}+ c^{\dagger})^2 } e^{-\frac{D}{2} \beta}
|0\rangle
\end{eqnarray}
for $D$-dimensional spacetime. Finally, using the Baker-Campbell-Haussdorf theorem again to write
\begin{equation}
e^{-\frac{i}{6\sqrt{2m\omega}}(a+a^{\dagger}) \cdot k_{23}} = e^{-\frac{i}{6\sqrt{2m\omega}}
a^{\dagger} \cdot k_{23}} e^{-\frac{i}{6\sqrt{2m\omega}} a \cdot k_{23}} e^{\frac{1}{2}
(-\frac{i}{6\sqrt{2m\omega}})^2 k_{23}^2}, \quad etc.,
\end{equation}
we find the 3-state interaction for the above massless state as
\begin{eqnarray}\label{A3result}
{\cal A}^{(0)}_3 & = & (\frac{1}{2m\omega})^{3/2} e^{-\frac{D}{2}\beta} \langle 0 | (\epsilon_1 \cdot a)
(\epsilon_2 \cdot b) (\epsilon_3 \cdot c) \nonumber \\ && \quad \quad \quad \quad \quad \quad \quad
\times e^{-\frac{i}{6\sqrt{2m\omega}}
[(a+a^{\dagger}) \cdot k_{23} +
(b+b^{\dagger}) \cdot k_{31} + (c+c^{\dagger}) \cdot k_{12}]}
e^{\alpha \frac{i}{6} (a^{\dagger}
+ b^{\dagger}+ c^{\dagger})^2 } |0 \rangle  \nonumber \\
& = & \kappa \{ \frac{-\frac{i}{6\sqrt{2m\omega}}\alpha}{3}
i [ (\epsilon_1 \
\cdot \epsilon_2) (\epsilon_3 \cdot k_{12}) + permutations ] \nonumber \\ && \quad \quad \quad \quad \quad
\quad + (-\frac{i}{6\sqrt{2m\omega}})^3 (\epsilon_1 \cdot k_{23}) (\epsilon_2 \cdot k_{31}) (\epsilon_3 \cdot k_{12})
\}
\end{eqnarray}
where
\begin{eqnarray}
\kappa &=& (\frac{1}{2m\omega})^{3/2} e^{-\frac{D}{2}\beta} e^{\frac{1}{2} (-\frac{i}{6\sqrt{2m\omega}})^2
(k_{23}^2 + k_{31}^2 + k_{12}^2)} \nonumber \\ &=& (\frac{1}{2m\omega})^{3/2} e^{-\frac{D}{2}\beta}
e^{-\frac{1}{48m\omega} (k_1^2 + k_2^2 + k_3^2)}
\end{eqnarray}
The result is the very same as the usual YM and $F^3$ 3-point interactions as obtained from bosonic open
string field theory in Feynman-Siegel gauge, except for the different ratio between the coefficients of the $F^2$ term and the
$F^3$ term. Also, (\ref{A3result}) has nonlocal coupling factors $\kappa$ as in bosonic
open string field theory.

Also, instead of using the operators of harmonic oscillators, direct Gaussian
integration gives exactly same result as above for the diagram in Fig.~\ref{fig2}. (\ref{A3}) can be
written as
\begin{equation}\label{A3integral}
{\cal A}^{(0)}_3 = \int d^D x d^D y d^D z \langle 0 | V_1 V_2 V_3 | x, y, z \rangle
\langle x,y,z | \widehat{F} |0 \rangle
\end{equation}
Substituting (\ref{externstate}) into it and integrating out all $x_1$, $x_2$, $y_1$, $y_2$, $z_1$ and
$z_2$, we get the same 3-point vertex for gauge bosons as (\ref{A3result}).

\begin{figure}[ht]
\begin{center}
\includegraphics[scale=0.5]{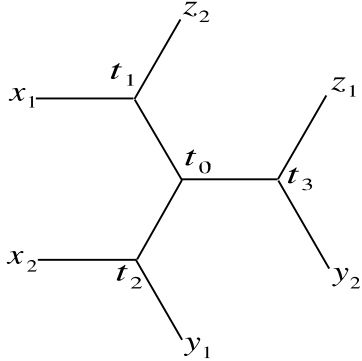}
\end{center}
\caption{\label{fig3}  The interaction lattice of order $\lambda^4$ with
the vertex given by $\widehat{G}$.}
\end{figure}

Another 3-string lattice vertex is shown in Fig.~\ref{fig3}, which is order $\lambda^4$ in the lattice coupling.
Then the 3-point amplitude is the same as in (\ref{A3}) with different 3-state vertex
\begin{eqnarray}\label{V31}
\widehat{G} &=& \int d^D X d^D Y d^D Z d^D t_0 d^D t_1 d^D t_2 d^D t_3
e^{-\frac{1}{2} [(t_1-t_0)^2+ (t_2-t_0)^2+ (t_3-t_0)^2] } \nonumber \\
& & \times e^{-\frac{1}{2} [(x_1-t_1)^2 + (x_2-t_2)^2 + (y_1 -
t_2)^2 + (y_2 - t_3)^2 + (z_1 - t_3)^2 + (z_2 - t_1)^2]}
e^{\frac{i}{2} (k_1
\cdot X + k_2 \cdot Y + k_3 \cdot Z)} \nonumber \\
& = & e^{-\frac{1}{10} (x^2+y^2+z^2) -\frac{1}{20} (x+y+z)^2 -
\frac{i}{10}(x \cdot k_{23} + y \cdot k_{31} + z \cdot k_{12}) }
\end{eqnarray}
With the external massless state of section \ref{5}, the 3-state interaction is
\begin{equation}\label{A3_1}
{\cal A}^{(1)}_3 = \lambda^4 (\langle 0|_1 \otimes
\langle 0 |_2 \otimes \langle 0 |_3 ) | V_1 V_2 V_3 \widehat{G} |0 \rangle
\end{equation}
In the operator formalism, the computation of $ {\cal A}^{(1)}_3$ is a little trickier.
Introduce three new variables through an orthogonal rotation:
\begin{eqnarray}
&& x' = \frac{1}{\sqrt{3}} (x+y+z) \nonumber \\
&& y' = \frac{1}{\sqrt{2}} (x-y) \\
&& z' = \frac{1}{\sqrt{6}} (x+y-2z) \nonumber
\end{eqnarray}
Thus
\begin{eqnarray}
e^{-\frac{1}{10} (x^2+y^2+z^2) -\frac{1}{20} (x+y+z)^2} = e^{-\frac{1}{4} x'^2
-\frac{1}{10} (y'^2+z'^2)}
\end{eqnarray}
Also, 3 pairs of new annihilation (creation) operators
\begin{eqnarray}
a' = \frac{1}{\sqrt{3}} (a+b+c) &,& a'^{\dagger} = \frac{1}{\sqrt{3}} (a^{\dagger}+b^{\dagger}+c^{\dagger})\nonumber \\
b' = \frac{1}{\sqrt{2}} (a-b) &,& b'^{\dagger} = \frac{1}{\sqrt{2}} (a^{\dagger}-b^{\dagger})\\
c' = \frac{1}{\sqrt{6}} (a+b-2c) &,& c'^{\dagger} = \frac{1}{\sqrt{6}} (a^{\dagger}+b^{\dagger}-2 c^{\dagger}) \nonumber
\end{eqnarray}
are introduced, which are independent of each other because
$[a' , b'^{\dagger}] = [a' , c'^{\dagger}]=0$, etc. Then
\begin{eqnarray}
e^{-\frac{1}{4} x'^2} = e^{-\frac{1}{4} \frac{1}{2m\omega} (a'^2 + a'^{\dagger 2}
+ \{a' , a'^{\dagger} \}) } \nonumber \\
e^{-\frac{1}{10} y'^2} = e^{-\frac{1}{10} \frac{1}{2m\omega} (b'^2 + b'^{\dagger 2}
+ \{b' , b'^{\dagger} \}) } \nonumber \\
e^{-\frac{1}{10} x'^2} = e^{-\frac{1}{10} \frac{1}{2m\omega} (c'^2 + c'^{\dagger 2}
+ \{c' , c'^{\dagger} \}) }
\end{eqnarray}
The ingredients of each exponent satisfy the commutation relations of raising and lowering operators for SU(1,1),
We use the same representation as in (\ref{split1}):
\begin{eqnarray}
& \frac{i}{2} (a')^2 \rightarrow \left(
\begin{array}{cc}
0 & 1 \\
0 & 0 \\
\end{array}
\right), \quad
\frac{i}{2} (a'^{\dagger} )^2 \rightarrow \left(
\begin{array}{cc}
0 & 0 \\
1 & 0 \\
\end{array}
\right) , \nonumber \\
&-\frac{1}{2} \{a', a'^{\dagger} \} \rightarrow
\left(
\begin{array}{cc}
1 & 0 \\
0 & -1 \\
\end{array}
\right)
\end{eqnarray}
and, with the same procedure, find
\begin{eqnarray}
e^{-\frac{1}{4} x'^2} =  e^{-\frac{1}{4} \frac{1}{2m\omega} (a'^2 + a'^{\dagger 2}
+ \{a' , a'^{\dagger} \}) } =
e^{\alpha_1 \frac{i}{2} a'^{\dagger 2}} e^{-\alpha_2 \frac{1}{2} \{a', a'^{\dagger} \}}
e^{\alpha_1 \frac{i}{2} a'^2}
\end{eqnarray}
Similarly,
\begin{eqnarray}
e^{-\frac{1}{10} y'^2} =  e^{-\frac{1}{10} \frac{1}{2m\omega} (b'^2 + b'^{\dagger 2}
+ \{b' , b'^{\dagger} \}) } =
e^{\beta_1 \frac{i}{2} b'^{\dagger 2}} e^{-\beta_2 \frac{1}{2} \{b', b'^{\dagger} \}}
e^{\beta_1 \frac{i}{2} b'^2}     \\
e^{-\frac{1}{10} z'^2} =  e^{-\frac{1}{10} \frac{1}{2m\omega} (c'^2 + c'^{\dagger 2}
+ \{c' , c'^{\dagger} \}) } =
e^{\beta_1 \frac{i}{2} c'^{\dagger 2}} e^{-\beta_2 \frac{1}{2} \{c', c'^{\dagger} \}}
e^{\beta_1 \frac{i}{2} c'^2}
\end{eqnarray}
Here $\alpha_1, \alpha_2, \beta_1, \beta_2$ are defined as:
\begin{eqnarray}
\alpha_1 = \frac{i}{1+ 2(2m\omega)} &,& \alpha_2 = \ln [1+ \frac{1}{2(2m\omega)}] \nonumber \\
\beta_1 = \frac{i}{1+ 5(2m\omega)} &,& \beta_2 = \ln [1+ \frac{1}{5(2m\omega)}]
\end{eqnarray}
Obviously, annihilation operators $a',b',c'$ also annihilate the vacuum $|0 \rangle$ and
\begin{eqnarray} & e^{-\frac{1}{10} (x^2+y^2+z^2) -\frac{1}{20} (x+y+z)^2} | 0 \rangle = C
e^{i \frac{\alpha_1}{6} (a+b+c)^2 +
i\frac{\beta_1}{3} (a^{\dagger 2} + b^{\dagger 2} + c^{\dagger 2} - a^{\dagger} \cdot b^{\dagger}
- b^{\dagger} \cdot c^{\dagger} -c^{\dagger} \cdot a^{\dagger}} | 0 \rangle \\
& C = e^{-\frac{\alpha_2}{2} D -\frac{\beta_2}{2} D -\frac{\beta_2}{2} D} \nonumber
\end{eqnarray}
Finally, using the Baker-Campbell-Haussdorf theorem directly, up to a constant,
\begin{eqnarray}\label{A31result}
&& {\cal A}^{(1)}_3 = \lambda^4 \langle 0 | V_1 V_2 V_3 \widehat{G} |0 \rangle \nonumber \\ & \propto & \kappa'
\{ \frac{4+40m\omega}{1+4m\omega}[ (\epsilon_1 \cdot \epsilon_2) (\epsilon_3 \cdot k_{12}) + permutations ]
+ (\epsilon_1 \cdot k_{23}) (\epsilon_2 \cdot k_{31}) (\epsilon_3 \cdot k_{12}) \}  \nonumber \\
\end{eqnarray}
with
\begin{equation}
\kappa' = i \lambda^4 e^{-\frac{3}{40} \frac{1}{1+10m\omega}
(k_1^2 + k_2^2 + k_3^2) }
\end{equation}
The exponents of $k_i^2$'s in $\kappa'$ will vanish if it is on-shell but will make
coupling factors nonlocal off-shell. Again, this result can be obtained by Gaussian integration
 directly, as in (\ref{A3integral}). We won't go through the details.

It is easy to notice that both vertices $\widehat{F}$ in (\ref{V30}) and $\widehat{G}$
in (\ref{V31}) give some gauge-fixed interactions for the massless state constructed from
partons, as in Witten's bosonic open string field theory,
which will be discussed in the next section.
Their comparison will be interesting because it
will give another view of string field theory, from the lattice.

\section{Comparison to string field theory}\label{7}

In this section, we will compare the two 3-state vertices mentioned in the last section
and the 3-state coupling from them with those in SFT.
As we will notice, if all oscillator modes but the zeroth and first
are truncated, the structure of 3-state vertices
$\widehat{F}$ in (\ref{V30}) and $\widehat{G}$ in (\ref{V31}) seem similar to the
3-string vertex from Witten's interaction in SFT, except for different
coefficients.

In above sections, the scale of the lattice was set to 1, which leads to the slope
of the Regge trajectory ${1\over 4\omega}$. So before comparing with string field
theory, we have to restore the scale of the lattice to match the slope with the Regge
slope from usual string theory (or string field theory).

We use the lattice actions (\ref{action}), with the lattice scale $\tilde{\alpha}'$.
The calculations in previous sections are unchanged except for rescaling the momenta by
\be k_i \rightarrow \sqrt{\tilde{\alpha}'} k_i \ee
and renormalizing the lattice coupling by
\be \lambda \rightarrow \lambda' \ee
Then the real Regge trajectory is
\be \alpha(s) = -\2 D +{1\over \omega}
[\frac{\tilde{\alpha}'}{4} s +ln(\lambda'^2) ] \ee
with the slope $\frac{\tilde{\alpha}'}{4 \omega}$. Setting it to be the same as
the Regge slope from string theory, which is $\alpha'$, we need the lattice scale
$$\tilde{\alpha}' = 4 \omega \alpha'$$
The intercept condition will be the same as
(\ref{intercept}) but replacing $\lambda$ by $\lambda'$.

It is easy to see the propagator (\ref{2point}) for the gauge boson in the lattice string
still has a massless pole. Also, the lattice
rescaling did nothing to either 3-string vertex but change the scale of momenta and
so change the ratio between coefficients of $F^2$ terms and $F^3$ terms in 3-point
amplitudes.

In string field theory, the general 3-string interaction can be interpreted as
\be \langle h_1 [\vartheta_A] h_2 [\vartheta_b] h_3 [\vartheta_c] \rangle = (
\langle A |_1 \otimes
\langle B |_2 \otimes \langle C |_3 ) | V_{123} \rangle \ee
where $\vartheta_i$ is the vertex operator for each external state and
$h_i(z)$ is the conformal mapping from each string strip to the complex plane \cite{lpp}.
In Witten's theory, the strings couple by overlapping the right half of each string with the
left half of the next \cite{Witten}.
Because there is only one oscillator mode in our ladder approximation for the lattice, here only the
zeroth and first level oscillator modes will be considered in SFT aspect.
After truncating oscillator modes and ignoring ghost contributions (there is no
worldsheet gauge fixing on the lattice), the 3-string vertex in the oscillator approach is \cite{Gross-Jevicki-123}:
\bea\label{V3Witten}
| V_{123} \rangle = {\mathcal N} \delta^D (k_1 + k_2 + k_3) \exp \Bigl(
  -\frac{1}{2}\sum_{I, J = 1}^{3}
[a^{I}_{-1} N^{IJ}_{-1,-1} a^{J}_{-1} +
2 a^{I}_{-1} N^{IJ}_{-1,0} p^{J} \nonumber \\ +
p^{I} N^{IJ}_{00}  p^{J}] \Bigr) | 0 \rangle_1 \otimes
| 0 \rangle_2 \otimes | 0 \rangle_3
\eea
The Neumann coefficients $N^{IJ}_{mn}$ depend on the choice of the conformal mappings.
It was shown that the different conformal mappings correspond to different formulations of string field theory that
are equivalent to each other. The most widely used open string field theory is Witten's theory.
The action in Witten's open string field theory is
\begin{equation}
S = + \frac{1}{2}\langle V_2 | \Psi, Q \Psi \rangle
  + \frac{g}{3}   \langle V_3 | \Psi, \Psi, \Psi \rangle\,.
\end{equation}
in which the first part gives the free term and the second part gives the interactions.
The Neumann coefficients read
\be N^{11}_{-1,-1} = N^{22}_{-1,-1} = N^{33}_{-1,-1} = \frac{5}{27} \ee
\be N^{12}_{-1,-1} = N^{23}_{-1,-1} = N^{31}_{-1,-1} = -\frac{16}{27} \ee
\be N^{12}_{-1,0} = - N^{13}_{-1,0} = N^{23}_{-1,0} = - N^{21}_{-1,0}= N^{31}_{-1,0} = - N^{32}_{-1,0} = \frac{2 \sqrt{6\alpha'}}{9} \ee
\be N^{11}_{00} = N^{22}_{00} = N^{33}_{00} = \alpha' \ln (27/16) \ee
and zero for others \cite{Taylor}.
With the Feynman-Siegel gauge $b_0 = 0$, for tachyon and massless states and up to 3-point interactions, we will get the
gauge-fixed action from the gauge-invariant action
\bea\label{S_gi} S
= \frac{1}{2} [\nabla_{\mu} , \phi] [\nabla^{\mu} , \phi] - \phi^2
- F_{\mu \nu} F^{\mu \nu} + \frac{1}{3} \phi^3 + 2 \phi F_{\mu \nu} F^{\mu \nu}
- \frac{4}{3} F_{\mu}^{\nu} F_{\nu}^{\lambda} F_{\lambda}^{\mu}
\eea
with a particular gauge as discussed in \cite{fs}. Here we focus on only the massless state and set
\be\label{A-state} \langle \psi_i |  = \langle 0,k_i |_I A(k_i) \cdot a_1^I \ee
It gives the 3-point gauge interactions
\bea {\cal A}_3 = \frac{g}{3}   \langle V_3 | \Psi, \Psi, \Psi \rangle
\propto i g e^{-\frac{1}{2} N^{11}_{00} (k_1^2 + k_2^2 + k_3^2)}
\{ [ (A_1 \cdot A_2) (A_3 \cdot k_{12}) + permutations ] \nonumber \\
+ \frac{\alpha'}{2} (A_1 \cdot k_{23}) (A_2 \cdot k_{31}) (A_3 \cdot k_{12}) \}
\eea
after $\alpha'$ is restored.

From the previous section, in the oscillator approach, the 3-state vertices in (\ref{V30}) or
(\ref{V31}) give the same form as (\ref{V3Witten}) except for some different Neumann coefficients.

For (\ref{V30}),
\be N^{11}_{-1,-1} = N^{22}_{-1,-1} = N^{33}_{-1,-1} = \frac{1}{3} \frac{1}{1+2m\omega}
\ee
\be N^{12}_{-1,-1} = N^{23}_{-1,-1} = N^{31}_{-1,-1} = \frac{1}{3} \frac{1}{1+2m\omega} \ee
\be N^{12}_{-1,0} = - N^{13}_{-1,0} = N^{23}_{-1,0} = - N^{21}_{-1,0}= N^{31}_{-1,0} = - N^{32}_{-1,0}
= \frac{\sqrt{\tilde{\alpha}'}}{3\sqrt{2m}} \ee
\be N^{11}_{00} = N^{22}_{00} = N^{33}_{00} = \frac{\tilde{\alpha}'}{6m} \ee
and zero for others. This gives a 3-point interaction for massless bosons
as in
(\ref{A3result}) with the ratio of $F^3$ and $F^2$ coefficients
\be \tilde{\alpha}' \frac{1+2m\omega}{6m} = \alpha' \frac{1+2m\omega}{3m} \omega \ee
instead of the ratio $\alpha'/2$ from Witten's vettex.

For (\ref{V31}),
\be N^{11}_{-1,-1} = N^{22}_{-1,-1} = N^{33}_{-1,-1} = \frac{1}{3} \frac{1}{1+4m\omega}
+ \frac{2}{3} \frac{1}{1+10m\omega} \ee
\be N^{12}_{-1,-1} = N^{23}_{-1,-1} = N^{31}_{-1,-1} = \frac{1}{3} [\frac{1}{1+4m\omega}
- \frac{1}{1+10m\omega}] \ee
\be N^{12}_{-1,0} = - N^{13}_{-1,0} = N^{23}_{-1,0} = - N^{21}_{-1,0}= N^{31}_{-1,0} = - N^{32}_{-1,0}
=  \sqrt{\tilde{\alpha}'} \frac{\sqrt{2m} \omega}{1+ 10m\omega} \ee
\be N^{11}_{00} = N^{22}_{00} = N^{33}_{00} =  \frac{3}{5} \tilde{\alpha}'
\frac{\omega}{1+10m\omega} \ee
and zero for others. Again, the 3-point interaction for the massless state is the same as in (\ref{A31result})
with a ratio of $F^3$ and $F^2$ coefficients of
\be \tilde{\alpha}' \frac{1+4m\omega}{1+10m\omega} \omega = 4 \alpha' \frac{1+4m\omega}{1+10m\omega} \omega^2 \ee

As Witten's theory in Feynman-Siegel gauge, both (\ref{V30}) and (\ref{V31}) give gauge-fixed
3-point interactions with nonlocal $e^{\tau \bo}$ factors.
The mismatch of $F^2$ and $F^3$ coefficients may be due to the fact we only considered the two simplest
interacting lattice diagrams. In principle, all interaction diagrams should be summed, which may give the
same interaction as from usual string theory (on-shell) or Witten's string field theory (off-shell).
But it does show that the massless state given in the beginning of section \ref{5} can have the same interactions
as the usual YM field. So this will be an interesting start to view the YM gauge field as the bound state of an underlying
scalar field instead of as a fundamental field.

\begin{figure}[ht]
\begin{center}
\includegraphics[scale=0.75]{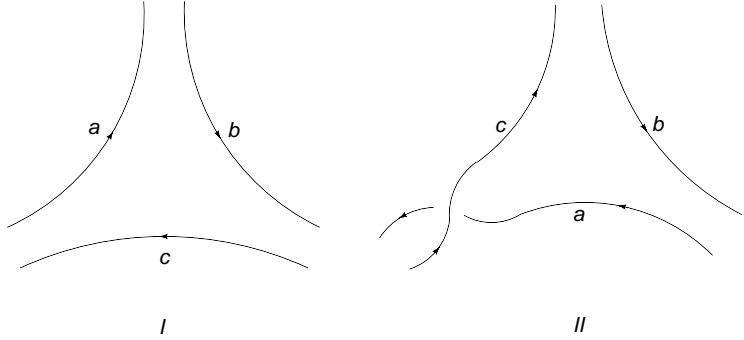}
\end{center}
\caption{\label{fig4} The twist symmetry in Witten's vertex: $I \propto Tr(T_a T_b T_c)$ ; $II
 \propto -Tr(T_c T_b T_a)$}
\end{figure}

Another similarity between vertices (\ref{V30}) or (\ref{V31}) and Witten's vertex is that they all have the same
symmetries. First, there is a cyclic symmetry under $I \rightarrow J, J \rightarrow K, K \rightarrow I$, which corresponds to cyclic symmetry of each interaction diagram. Second, there is a symmetry for Neumann coefficients under $I \leftrightarrow J , m \leftrightarrow n$. Finally, there is a twist symmetry under $N^{IJ}_{nm} = (-1)^{m+n} N^{JI}_{nm}$
associated with twisting of the lattices (strings). It is nontrivial and restricts the group structure of the gauge-fixed action. For the case here, we only consider the first excited
state $|\psi_i \rangle$ in (\ref{A-state}), which is a twist-odd 
state under the twist operator $\Omega$: $\Omega |\psi_i \rangle = -|\psi_i \rangle$. 
Then the twist invariance requires the gauge-fixed interaction to be proportional to the structure constants $f^{abc}$
because 
\be \Omega \langle \Psi_1, \Psi_2 * \Psi_3 \rangle = \langle (\Omega \Psi_1), (\Omega \Psi_3) 
* (\Omega \Psi_2) \rangle = - \langle \Psi_1, \Psi_3 * \Psi_2 \rangle 
\ee
as shown in Fig~\ref{fig4}.
There, diagram $I$ gives the term $\propto Tr(T_a T_b T_c)$ while diagram $II$ gives the term
$\propto -Tr(T_c T_b T_a)$ and their sum gives an interaction term $\propto f^{abc}$.
Because the gauge-invariant YM action can
always be written as a function of structure constants, the gauge condition in this case should
also be expressed in terms of $f^{abc}$, which excludes the Gervais-Neveu gauge.
These symmetries apply not only to massless states but also to
general states (but with the usual extra sign factors in the twist). Obviously, both Fig.~\ref{fig2} and  Fig.~\ref{fig3} are similar to the diagram
of joining three open strings in Witten's theory except they are on a discrete lattice while
Witten's vertex is on a continuous worldsheet.

Another 3-string vertex in SFT we will mention here is the CSV vertex, which is equivalent
to Witten's vertex on-shell.
Here we only review the coefficients for zero-modes and first excited modes:
\be N^{11}_{-1,-1} = N^{22}_{-1,-1} = N^{33}_{-1,-1}= 0 \ee
\be N^{12}_{-1,-1} = N^{23}_{-1,-1} = N^{31}_{-1,-1}= 1 \ee
\be N^{12}_{-1,0} =  N^{23}_{-1,0}= N^{31}_{-1,0} = \sqrt{2 \alpha'} \ee
\be N^{21}_{-1,0} =  N^{32}_{-1,0}= N^{13}_{-1,0} = 0 \ee
and all $N^{IJ}_{00}$ vanish.
Comparing to the above vertices, the CSV vertex lacks twist symmetry. So, as has
been shown previously, it corresponds to the well-known Gervais-Neveu gauge
without nonlocal coupling factors.

\section{Discussions}\label{8}

In this article, we started from the bosonic lattice string, and constructed the
massless state as a bound state of partons, and two simple lattice interaction diagrams.
Using such interaction
diagrams, we found interactions of those bound states similar to the usual YM gauge field.
The comparison of these 3-state vertices on the lattice with Witten's vertex on the continuous worldsheet
shows all of them have the same symmetries, especially twist symmetry, which is absent in the CSV vertex.
The twist symmetry restricted the gauge-fixed interaction to be proportional to the structure
constants of the gauge group, or equivalently, the interaction term of the gauge condition must be proportional to the structure
constants. That's the reason the Gervais-Neveu gauge can only be obtained from the CSV vertex. Anyway,
we show here the possibility to bind the scalars on the lattice to get the massless
vector state which behaves like the gauge field, i.e., the gauge field is no longer a fundamental
particle but a composite state in the field theory. This also provided a new view of the 3-string coupling
in Witten's bosonic open string field theory.

\section*{Acknowledgement}

This work is supported in part by National Science Foundation
Grant No.\ PHY-0354776.


\begin{thebibliography}{99}

\bibitem{Regge}
T. Regge, {\it Nuo. Cim.} {\bf 14} (1959) 951; {\bf 18} (1960) 947.

\bibitem{bethe} H.A. Bethe and E.E. Salpeter,  {\it Phys. Rev.} {\bf 84} (1951) 1232

\bibitem{random} M.R. Douglas and S.H. Shenker, {\it Nucl. Phys.} {\bf B335} (1990) 635;\\
D.J. Gross and A.A. Migdal, {\it Phys. Rev. Lett.} {\bf 64} (1990) 127;\\
E. Br\'ezin and V.A. Kazakov, {\it Phys. Lett.} {\bf 236B} (1990) 144.

\bibitem{David}
F. David, {\it Nucl. Phys.} {\bf B257} [FS14] (1985) 543; \\
V.A. Kazakov, I.K. Kostov, and A.A. Migdal, {\it Phys. Lett.} {\bf 157B} (1985) 295;\\
J. Ambj\o rn, B. Durhuus, and J. Fr\"ohlich, {\it Nucl. Phys.} {\bf B257} (1985) 433.

\bibitem{ladders}
T. Biswas, M. Grisaru and W. Siegel, {\it Nucl.Phys.} {\bf B708} 2005, 317.

\bibitem{eden}
R.J. Eden, P.V. Landshoff, D.I. Olive and J.C. Polkinghorne,
{\it The Analytic S-Matrix} (Cambridge University Press, 1966);\\
M. Froissart, {\it Phys. Rev.} {\bf 123} (1961) 1053;\\
V.N. Gribov, {\it JETP} {\bf 14} (1962) 1395; \\
V.N. Gribov and I. Ya. Pomeranchuk, {\it Phys. Rev. Lett.} {\bf 9}  (1962)
238; \\
D. Amati, S. Fubini and A. Stangellini, {\it Nuovo Cimento} {\bf 26} (1962) 896; \\
B.W. Lee and R.F. Sawyer, {\it Phys. Rev.} {\bf 127} (1962) 2266.

\bibitem{federbush}
P.G. Federbush and M.T. Grisaru, {\it Ann. Phys.} {\bf 22} (1963) 263,
299;\\
J.C. Polkinghorne, {\it J. Math. Phys.} {\bf 4} (1963) 503.

\bibitem{Hooft}
G. 't Hooft, {\it Nucl. Phys.} {\bf B72} (1974) 461.

\bibitem{W. Siegel1}
W. Siegel, \hhref{9601002}, {\it Int. J. Mod. Phys. A} {\bf 13} (1998) 381.

\bibitem{susskind}
L. Susskind, {\it Phys.Rev.Lett.} {\bf 23} (1969) 545.

\bibitem{lpp}
A.\ Leclair, M.\ E.\ Peskin, and C.\ R.\ Preitschopf,
{\it Nucl. Phys.} {\bf B317} (1989) 411; 464.

\bibitem{Witten}
E. Witten, {\it Nucl. Phys.} {\bf B268} (1986) 253.

\bibitem{Gross-Jevicki-123}
D.\ J.\ Gross and A.\ Jevicki, \NP {\bf B283} (1987), 1; \NP {\bf B287}
(1987), 225; \NP {\bf B293} (1987) 29;\\
E. Cremmer, C.B. Thorn, and A. Schwimmer, {\it Phys. Lett.} {\bf 179B} (1986) 57.



\bibitem{Taylor}
W.\ Taylor and B.\ Zwiebach, \hhref{0311017},
{\it Boulder 2001, Strings, branes and extra dimensions} (2003) 641.

\bibitem{fs}
H. Feng and W. Siegel, \hhref{0611307}.


\end{thebibliography}
\end{document}